\documentclass[
    reprint,
    preprintnumbers,
    superscriptaddress,
    nofootinbib,
     amsmath,amssymb,
     aps,
     prd,
    floatfix,
    longbibliography]{revtex4-2}
    
    \usepackage{graphicx}
    \usepackage[caption=false]{subfig}
    \usepackage[usenames,dvipsnames]{xcolor} 
    \usepackage{pgfplots}
    \usepackage[utf8]{inputenc}
    \usepackage{color}
    \usepackage{hyperref}
    \usepackage[normalem]{ulem} 
    \usepackage{physics}
    \usepackage{enumitem}
    \usepackage{comment}
    \usepackage{bm}
    \usepackage{aas_macros}
    \usepackage[capitalise]{cleveref}
    \usepackage{caption}

\hypersetup{
  breaklinks = true,
  colorlinks   = true, 
  urlcolor     = Blue, 
  linkcolor    = Blue, 
  citecolor   = Blue 
}
    \allowdisplaybreaks[1]

\usepackage{float}

\begin{document}

\title{Dispersion in the Hubble-Lema\^{i}tre constant measurements from gravitational clustering}

\author{Swati Gavas}
\email{swatigavas47@gmail.com}
\affiliation{Indian Institute of Science Education and Research (Mohali), Sector 81, SAS Nagar, Punjab 140306, India}
\author{J S Bagla}
\email{jasjeet@iisermohali.ac.in}
\affiliation{Indian Institute of Science Education and Research (Mohali), Sector 81, SAS Nagar, Punjab 140306, India}
\affiliation{National Centre for Radio Astrophysics, Tata Institute of Fundamental Research, Ganeshkhind, Pune 411007, India}
\author{Nishikanta Khandai}
\email{nkhandai@niser.ac.in}
\affiliation{School of Physical Sciences, National Institute of Science Education and Research, Jatni 752050, India}
\affiliation{Homi Bhabha National Institute, Training School Complex, Anushaktinagar, Mumbai 400094, India}

\begin{abstract}
    Measurements of the Hubble-Lema\^{i}tre constant ($H_0$) require us to estimate the distance and recession velocity of galaxies independently.   
    Gravitational clustering that leads to the formation of galaxies and the large scale structure leaves its imprints in the form of peculiar velocities of galaxies.  
    In general, it is not possible to disentangle the peculiar velocity component from the recession velocities of galaxies, and this introduces an uncertainty in the determination of $H_0$.  
    We use cosmological N-body simulations to quantify the impact of peculiar velocities on the estimation of $H_0$.  
    We consider observers to be located in dark matter halos and target galaxies to be distributed amongst dark matter halos.  
    We compute the distribution of the estimated value of $H_0$ across all such observers in the simulation, and we study the distribution as a function of distance from the observer.  
    We find that the dispersion of this distribution is large at small scales, and it diminishes as we go to large separations, reaching the level of the quoted statistical error in Planck and SH0ES measurements well beyond $\sim$135 Mpc/h and $\sim$220 Mpc/h, respectively.   
    Measurements at smaller scales are susceptible to errors arising from peculiar motions, and this error can propagate to measurements at larger scales in the distance ladder.  
    Notably, we observe a negative correlation between the local over-density around an observer and the deviation of the local and the global value of $H_0$.  
    We show that deviations more significant than $5$\% of the global values can be encountered frequently at scales of up to $40$~Mpc/h, and this is considerably larger than the statistical errors on local estimates.
    We also analyse the cumulative effect of such errors on mock measurements of $H_0$ as measured from Milky Way-sized halos. 
    We find that this error is sensitive to the lowest distance at which we use measurements.  
    The distribution of $H_0$ in mock measurements has a large tail, and deviations of a few percent from the global value cannot be ruled out.
\end{abstract}

\maketitle

\section{Introduction}

Hubble-Lemaître constant ($H_0$) is a key parameter in relativistic cosmology.
It encapsulates the rate of expansion of the Universe at present \cite{freedman2010,tully2023, verde2023}. 
Being a dimensional quantity, it is also a measure of the age of the Universe in combination with other cosmological parameters.
Measurements of $H_0$ at scales of a few to a few tens of megaparsecs are done using standard candles like Cepheid variables.
A number of distance indicators are used to extend these measurements to larger distances.
Measurements of $H_0$ are also made using time delay measurements in gravitational lensed systems and CMBR anisotropies \cite{brieden2023}. 
There is, at present, a strong tension between the measured values of $H_0$ using local and distant sources \cite{plank2020,riess2021,riess2022,brout2022}. 
This has been given the name Hubble tension \cite{shah2021,divalentino2021,verde2023}.   
Many possible extensions of the standard cosmological model have been proposed to account for this discrepancy, though there appears to be no single solution that accounts for all observables \cite{efstathiou2021,vagnozzi2023}.  
A number of studies have also been carried out to determine whether the contribution of systematic errors can help alleviate tension
\cite{mortsell2022,huterer2023,majaess2024a,2024arXiv240605047S,2024arXiv240605049C}.

Local inhomogeneities, such as supervoids and peculiar velocities, significantly influence the measurement of the $H_0$ \cite{2017MNRAS.471.4946W,2023CQGra..40i4001K,2023MNRAS.524.1885W,ruiz-lapuente2023,huang2024a,kalbouneh2024,2024arXiv240520409C,2024JCAP...01..071G}. 
Peculiar velocities, resulting from gravitational interactions between nearby galaxies, distort the smooth Hubble flow, affecting the observed recession velocities and subsequent distance measurements \cite{maartens2023}. Despite continuous efforts to improve distance measurement techniques and methodologies, discrepancies persist \cite{anand2024,riess2024,verde2023}, highlighting the importance of robust statistical analyses.

In this work, we study the effect of inhomogeneities and gravitational clustering on the measurements of $H_0$.
Our approach is inspired by \cite{turner1992}, where we use cosmological N-Body simulations to estimate the effect of peculiar velocities on the measurement of $H_0$.  
Similar studies have been carried out by others \cite{2014JCAP...10..028O,2014MNRAS.438.1805W,2017JCAP...03..022O}.  
In \cite{2014MNRAS.438.1805W}, it was found that observers in random dark matter halos typically measure lower expansion rates than those in voids, with differences more pronounced in halo rest frames compared to the CMB frame. \cite{2014JCAP...10..028O} explored how $H_0$ measurements are influenced by factors such as survey depth, size, sky coverage, and observer position. In \cite{2017JCAP...03..022O}, results from linear perturbation theory were compared with non-linear velocity power spectra derived from N-body simulations, which typically show smaller variances. Although none of these approaches produces a variance large enough to explain the $H_0$ discrepancy fully, the differences are significant given the precision of local $H_0$ measurements.
We use N-body simulations and consider galaxies to be embedded in dark matter halos.
Considering observers to be located in galaxies, we compute the {\it measured} $H_0$ for each galaxy pair by considering the peculiar velocities.
This allows us to construct and analyse the distribution for $H_0$ across different relative distance intervals for all observers.
The manuscript is structured as follows: In \S 2, we discuss the methodology to estimate local $H_0$ measurements. \S 3 outlines our simulations. In \S 4, we present results derived from our data analysis. Finally, we conclude our findings and discuss future directions in \S 5.

\section{Methodology}\label{methodology}

Assuming galaxies reside within dark matter halos, we use halos as a representative for both observers and observed galaxies. 
We use the distribution of halos identified in N-body simulations to create a group of hypothetical observers positioned on the halos.
We assign a single galaxy or observer for each halo: this is clearly a simplification, especially for massive halos, and leads to an
under-estimate the effect of peculiar velocities as we ignore the contribution of the intra-halo velocity distribution and the halo occupation distribution. 
We then derive the local Hubble constant ($H_L$) for observers in their neighbourhood volume by considering distances and peculiar
velocities of halos within the volume.
As a result, we get a set of local Hubble constant measurements, creating a distribution of measured values for different observers.
This distribution is then compared with the global Hubble constant ($H_G$) value used in the simulation model.
This approach closely follows the methodology used by \cite{turner1992}.

The contribution of cosmological expansion as well as peculiar velocities is combined to obtain redshifts that are then used to compute the local 
value of Hubble's constant for a given observer.
To set the notation, we define $\mathbf{r}_i$ as the proper coordinate of the $i^{\text{th}}$ halo.  This is related to the comoving coordinate as: $\mathbf{r}_i = a(t) \mathbf{x}_i$.  Hence, the velocities in the two coordinates are related as: $\mathbf{V}_i = \dot{a} \mathbf{x}_i + a(t) \dot{\mathbf{x}}_i = H \mathbf{r}_i + a(t) \dot{\mathbf{x}}_i $.  For the present epoch where we take $a(t_0) = 1$, we can write this as: $\mathbf{V}_i = H_0 \mathbf{r}_i + \dot{\mathbf{x}}_i$.  
The "observed" Hubble's constant by an observer at the origin is then $\mathbf{V}_i.\mathbf{r}_i / r_i^2$.  
For a pair of halos $i$ and $k$, the measured Hubble's constant is given as:
\begin{equation}
H_{ik} = H_0  + \frac{\left(\mathbf{x}_i - \mathbf{x}_k\right).\left(\dot{\mathbf{x}}_i - \dot{\mathbf{x}}_k\right)}{x_{ik}^2}.
\end{equation}
Here, $x_{ik} = |\left(\mathbf{x}_i - \mathbf{x}_k\right)|$, $\mathbf{x}$ represents the comoving coordinates and $\dot{\mathbf{x}}$ is the peculiar velocity.  
Note that we have $a_0=1$ where $a_0$ is the present value of the scale factor. 
The measured Hubble's constant for the $k^{\text{th}}$ galaxy is the sum of this quantity over a set of galaxies (halos).
The deviation of the measured Hubble's constant by the $k^{\text{th}}$ galaxy is then: 
\begin{equation}
    \label{eqn:delta_k}
\delta_{H\, k} = \frac{1}{N} \sum_{i = 1}^{N} \frac{{\bf \hat{x}}_{ik} {\bf \cdot
    (\dot{\mathbf{x}}_i - \dot{\mathbf{x}}_k)}}{H_G |{\bf x_i} - {\bf x_k}|}.
\end{equation}
The distribution of deviations from the reference value is described using the variable $\delta_H$ defined below in equation \ref{eqn:delta}.
$\delta_H$ is a fractional deviation of the local Hubble value from the global Hubble value
\begin{equation}
    \label{eqn:delta}
    \delta_H \equiv \frac{H_L - H_G}{H_G}.
\end{equation}
Here, $N$ represents the total number of halos within volume $V$, ${\bf \hat{x}}_{ik}$ corresponds to the unit vector pointing in the direction of the $i^{th}$ halo from the observer's location.
Averaging over all angles clearly reduces the dispersion in values, and hence $\delta_k$ is expected to be smaller than the corresponding estimation done in some patches in the sky for an observer. 

The size and the distance to the local volume affect the distribution $\delta$.
We define this local volume as a spherical shell centred on the observer and use multiple concentric shells to explore how $\delta$
varies with distance from the observer.
The lower radius ($r_{\text{min}}$) and upper radius ($r_{\text{max}}$) of the used volumes are displayed in the sixth column of table \ref{tab:sim}.  

While the estimation of Hubble's constant in shells is useful to understand the impact of peculiar velocities as a function of distance, real measurements are done over a wide range of scales using multiple distance indicators and cross-calibration. 
To mimic this, we also conduct mock observations over a wide range of shells to understand the impact of peculiar velocities on the measurement of $H_0$ in an ideal scenario.  
We restrict this part of the study to observers in Milky Way-size halos by mass.

\section{Simulations} \label{sec:simulations}

\begin{table*}
  \fontsize{9pt}{9pt}\selectfont
  \begin{center}
    \begingroup
    \setlength{\tabcolsep}{9pt}
    \renewcommand*{\arraystretch}{1.5}
    
    \begin{tabular}{c c c c c c c}
      \hline 
      $L_{\text{box}}$ & $N_{\text{part}}$ & $m_{\text{part}}$ & $\epsilon$  & ${N_{\text{halos}}}$ &  ${m_{\text{low}}}$ & Spherical shells: ($r_{\text{min}}$-$r_{\text{max}}$)\\
      (Mpc/h) & &  $(M_{\odot} $/h)& (Kpc/h) & & $(M_{\odot}$/h) &  (Mpc/h)\\ [0.5ex] 
      \hline 
      150 & $1024^3$ & $2.71 \times 10^8$ & 5 & 1,081,171 & $8.68 \times 10^9$ & (15-20), (20-25), (25-30), \\ & & & & & & (30-35), (35-40), (40-45),\\ & & & & & & (45-50), (50-55) \\ 
      \hline
      500 & $1024^3$ & $1.01 \times 10^{10}$  & 16.6 & 1,599,064 & $3.22 \times 10^{11}$ & (40-55), (55-70), (70-85),\\ & & & & & & (85-100), (100-115), (115-130),\\ & & & & & &  (130-145), (145-160) \\
      \hline
      1000 & $1024^3$ & $8.04 \times 10^{10}$ & 32.2 & 1,973,556 & $2.57 \times 10^{12}$ & (135-160) (160-185), (185-210), \\ & & & & & &(210-235),   (235-260), (260-285),\\ & & & & & & (285-310), (310-335)\\
      \hline
      
    \end{tabular}
    \endgroup
    \caption{\label{tab:sim}{\bf Simulation details:} Column 1: Side length of cubical simulation box, Column 2: Number of particles in the 
      simulation, Column 3: Mass of a single particle, Column 4:
      Softening length, Column 5: Total number of halos in each
      simulation (also equal to the number of individual observers), Column 6: Lowest Halo mass,
      Column 7: Spherical shells used as a local volume.} 
  \end{center}
\end{table*}

For simulations, we adopt the latest Planck-18 cosmology \citep{plank2020} with the following parameters: $\Omega_{m0} =
0.31$, $\Omega_{\Lambda0} = 0.69$, $\Omega_{b0} = 0.049$, $H_0 = 67.66$, $\sigma_8 = 0.81$, and $n_s = 0.9665$. 
We analyze data using the simulation output at redshift $z=0$ for this work. 

We use three dark matter-only simulations with a different cubical comoving box size: $150$~Mpc/h, $500$~Mpc/h, and $1000$~Mpc/h.
The large scale structure evolved in the simulations has significant cosmic variance on the scales comparable to box size and is also affected by discreteness errors at scales comparable to the grid size.
Therefore, we use three box sizes to probe a sufficient range of distance scales with improved reliability. 

All three simulations use $1024^3$ particles and are subject to periodic boundary conditions.
To ensure the consistency of results, we maintain overlap across the simulations in the spherical shells used as local volumes.
For more detailed information on the simulations, we refer the reader to table \ref{tab:sim}. 

The simulations were carried out using the {\tt GADGET4} code \citep{springel2021}, and the halo catalogs were produced using its built-in {\tt SUBFIND} algorithm.
The halo finder identifies Friends-Of-Friends (FOF) groups within the given distribution and decomposes each detected object into substructures using the excursion set algorithm. A linking length of 0.2 times the grid length is used, and all other default parameters of the code are used. Here, we focus on isolated halos, disregarding any halo part of a larger group or cluster. Substructures or sub-halos are not considered distinct entities; thus, each halo has only one observer, and additional galaxies/observers are not assigned to subhalos. Note that this leads to an elimination of the finger of god effect, leading to an underestimate of the effect of peculiar velocities. Thus, our results are expected to underestimate the actual effect. We verify that including substructures, increases observers by $\sim$30\% and increases dispersion by a few percent.

One of the issues that can affect our results is the difference between a single time snapshot as compared to a light cone snapshot from simulations. We can estimate the variation of peculiar velocities at large scales using the Zel'dovich approximation \citep{Zel'dovich_1970, Shandarin_1989}.  This is especially pertinent within our approach where we do not consider sub-halos.  It can be shown that the overall variation is that peculiar velocities come down over time \citep{Lahav_1991}, and hence, for the largest separations used here, we underestimate the effect of peculiar velocities by a few percent.  Note that the peculiar velocities are underestimated by a few percent, hence the overall effect on the dispersion in Hubble constant measurements is expected to be much smaller and can be ignored at this level. Similarly, we disregard the impact of baryonic processes on peculiar velocities, as studies have shown that this effect is well below 1\% at the scales considered in our analysis (\cite{Dolag_2013,Hellwing_2016,Kuruvilla_2020}).


\section{Results}\label{sec:results}

\subsection{The \texorpdfstring{$\delta_H$}\ \ distributions}

\begin{figure*}
    \centering
    \includegraphics[width=.88\textwidth]{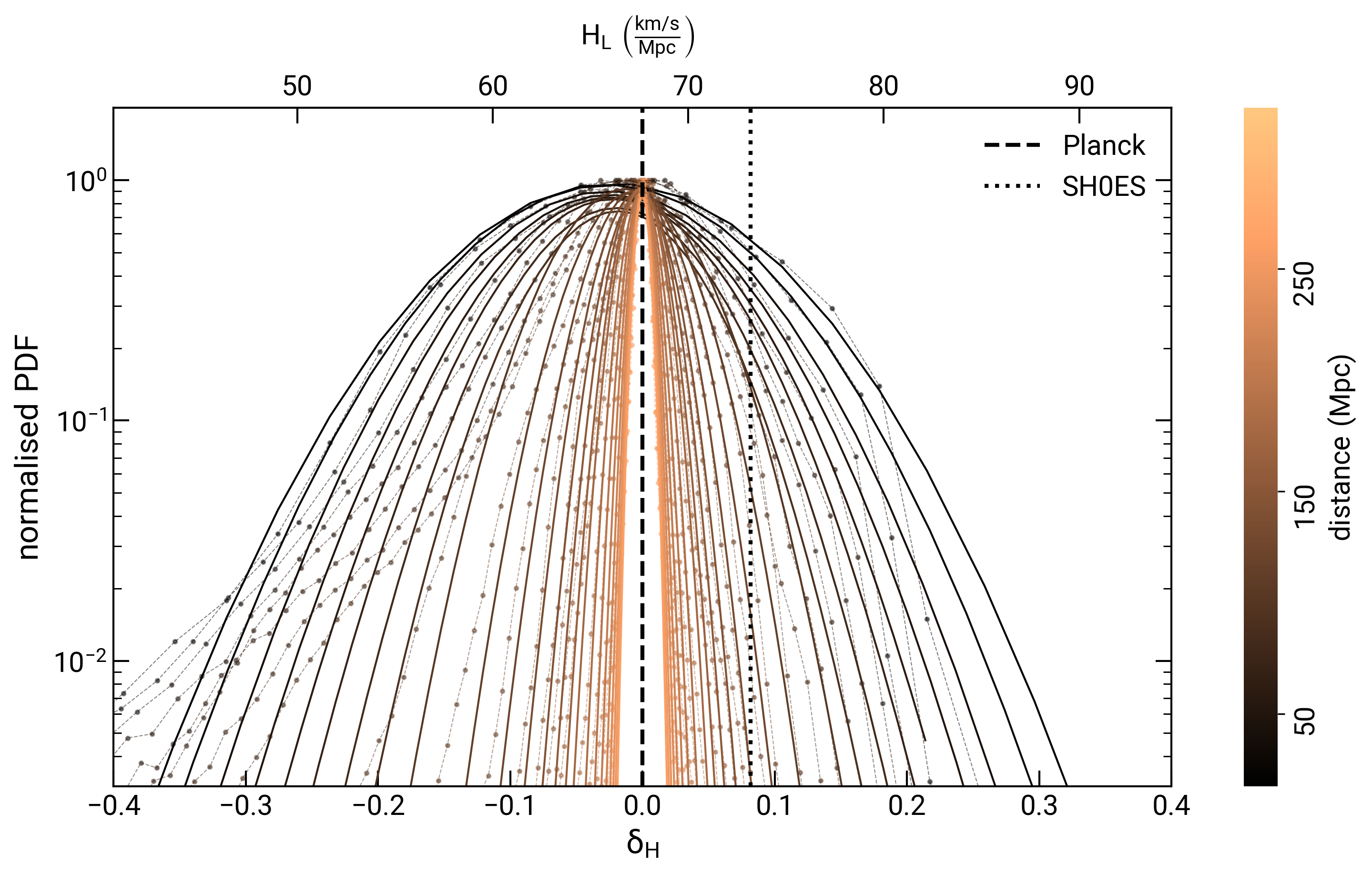}
    \caption{{\bf Probability distribution functions:} normalized PDFs of the fractional deviation within a spherical local volume in redshift space. The scatter points represent the PDF obtained from the simulation data, while the solid lines represent Gaussian functions characterized by the first two moments of the PDF. The colors of lines and scatter points correspond to the distance to the local volume from the observer, as shown in the adjacent colorbar.}
    \label{fig:pdf1}
\end{figure*}

The distribution of $\delta_H$ obtained using equations \ref{eqn:delta} and \ref{eqn:delta_k}, considering different spherical shells in the
three simulations, are illustrated in figures \ref{fig:pdf1}.
Here, the distribution is obtained by averaging over all halos with a pairwise distance in redshift space corresponding to the given bin.  
The figure has points corresponding to the probability distribution in each bin and the Gaussian corresponding to the first two moments of the PDF. 
The colors of lines and scatter points correspond to the distance of local volume from the observer, as shown in the adjacent colorbar.

We observe a significant dispersion in the local $H_0$ values at all probed distances, with a maximum deviation on negative(positive) sides varying from $\sim 50$\%($70$\%) at $40$ Mpc/h to $\sim 5$\%($6$\%) at $335$ Mpc/h. As we approach larger distances, the PDFs become narrower, and the peaks shift to zero due to the increasing contribution of the Hubble flow to velocities.

PDFs show heavy negative halves, which is more evident for small-volume shells. Differences in peak location between PDFs (scatter) and Gaussian functions (solid lines) exhibit deviations of varying magnitude. These point towards a significant non-Gaussianity of
the distributions.


\subsection{Statistics of distributions with distance}
\label{ssec:stat_with_d}

\begin{figure*}
    \centering
    \includegraphics[width=.88\textwidth]{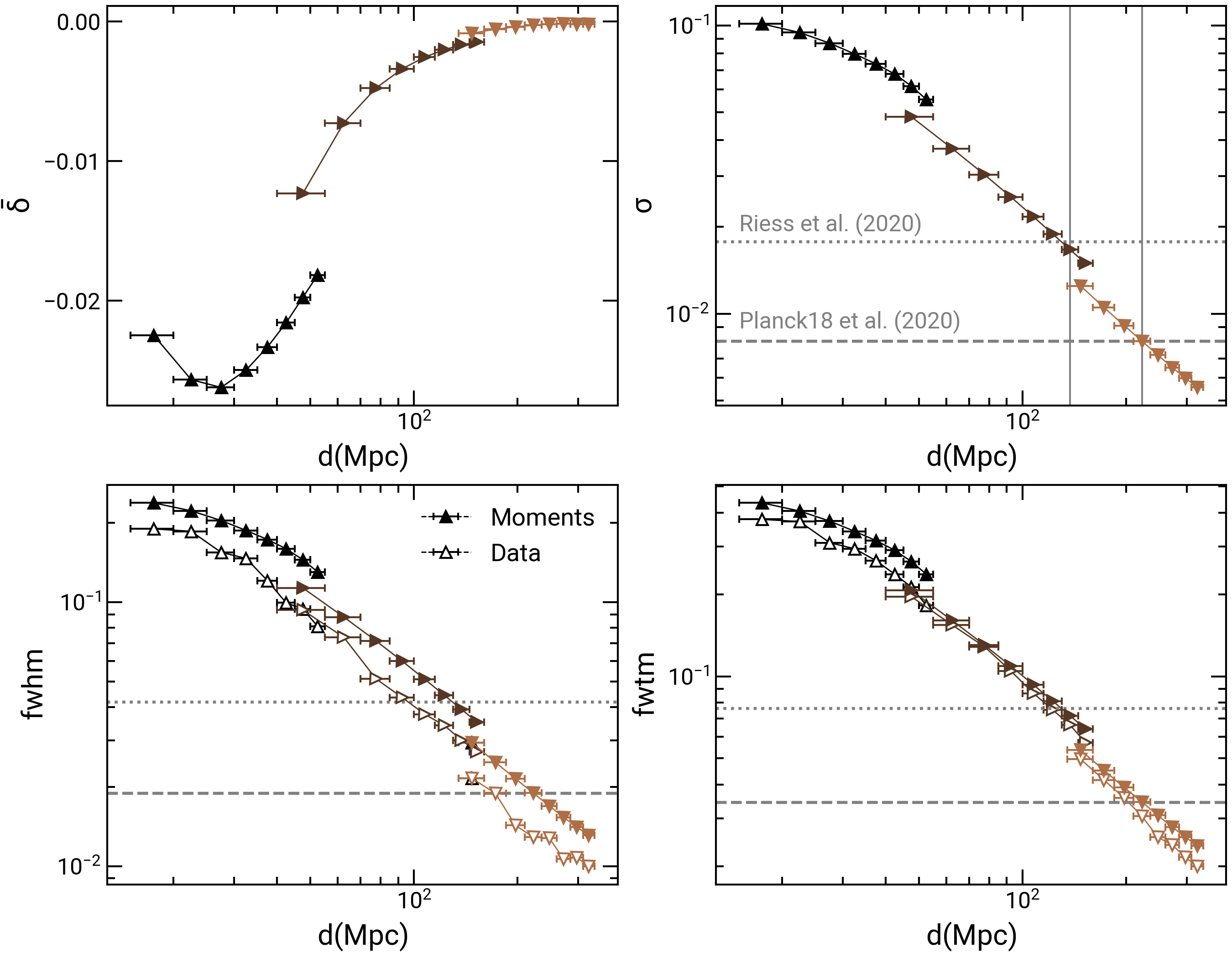}
    \caption{{\bf Statistics of distribution with distance:} Mean ($\bar{\delta}$), standard deviation ($\sigma$), fwhm, and fwtm as indicated on the y-axis with distance in the x-axis. Filled markers represent quantities derived from the moments of PDF, and empty markers represent quantities computed from the discrete PDFs. The marker orientations up, right, and down correspond to simulations with box sizes of $150$ Mpc/h, $500$ Mpc/h, and $1000$ Mpc/h, respectively. The dotted and dashed gray lines depict quantities reported by \protect\cite{riess2021} and \protect\cite{plank2020}.}
    \label{fig:stat_d}
\end{figure*}

We analyze the $\delta_H$ distributions by calculating various statistical quantities such as mean ($\bar{\delta}$), standard deviation ($\sigma$), full-width half maximum (fwhm), and full-width tenth maximum (fwtm). In figure \ref{fig:stat_d}, we present these quantities as a function of distance in separate panels. In the figure, filled markers represent moments and quantities derived from the moments of PDF assuming a Gaussian distribution, while empty markers represent quantities directly computed from the discrete PDFs. Up, right, and down orientations of triangular markers correspond to simulations with box sizes of $150$~Mpc/h, $500$~Mpc/h, and $1000$~Mpc/h, respectively. Additionally, the dotted and dashed grey lines depict quantities reported by \cite{riess2021} and \cite{plank2020}, providing a reference for comparison.

From the first panel of figure \ref{fig:stat_d}, we notice the mean approaches zero from the negative side with an increase in distance. This behaviour is expected as halos reside in denser environments than the average density of the Universe. As a result, the velocities of nearby halos are slowed, leading to an overall decrease in the measured value of $H_0$. This effect diminishes with an increase in the distance and size of the local volume.  A jump in the halo mass threshold causes the jump between the points from simulations of different sizes.

In the second panel of figure \ref{fig:stat_d}, we show the variation of $\sigma$, which consistently decreases as distance increases. The parameter $\sigma$, represents the expected statistical error in the measurement of the Hubble constant, originating from the varying positions of observers within the large-scale structure. It indicates that deviations from the global value, greater than $\sigma(d)$, in the measured Hubble constant $H_L$ at a given distance $d$, are less likely to occur.
As we investigate larger distances, this expected deviation diminishes, eventually leading to the convergence of the local ($H_L$) and global ($H_G$) Hubble constant values. The dotted and dashed gray lines in the plot represent the standard deviations quoted in observational measurements for local \citep{riess2021} and global \citep{plank2020} Hubble constant values, respectively. Given these factors and the existing errors in observational measurements, our analysis suggests a lower limit of $\sim 135$~Mpc/h for SH0ES and $\sim 220$~Mpc/h for Planck to get a robust convergence of these two Hubble constant values. 

Given the non-Gaussian nature of distributions, we use fwhm and fwtm in addition to mean and standard deviation for better description. Panels 3 and 4 of figure \ref{fig:stat_d} display the behavior of fwhm and fwtm. These plots reinforce our earlier conclusion regarding the lower limit for converging local and global Hubble constant values. Furthermore, they indicate that at a distance of $100$ Mpc/h, the probabilities (according to Gaussian) of encountering deviations exceeding 4\% and 10\% are $\sim$24\% and $\sim$3\%, respectively. These probabilities steeply decline as we move to distances beyond $100$ Mpc/h. Notably, the observed deviation of 8-9\% in observational measurements is rare in this context.

From the figure \ref{fig:stat_d}, we observe an offset in quantities as we transition across three simulations. The primary source of these offsets arises from the incomplete representation of the large-scale velocity field in simulations due to the finite box size. Smaller box sizes are more susceptible to this limitation, as the lack of large-scale information impacts peculiar velocities more significantly than the density field. The variation in the number of independent modes on small and large scales in simulations can be responsible for some of the contribution to the offsets in sigma between simulations.Additionally, finite mass resolution can be a secondary source of offsets across different simulations.


\subsection{Correlation with local density}
\label{ssec:corre}

\begin{figure*}
    \centering  
        \includegraphics[width=.35\textwidth]{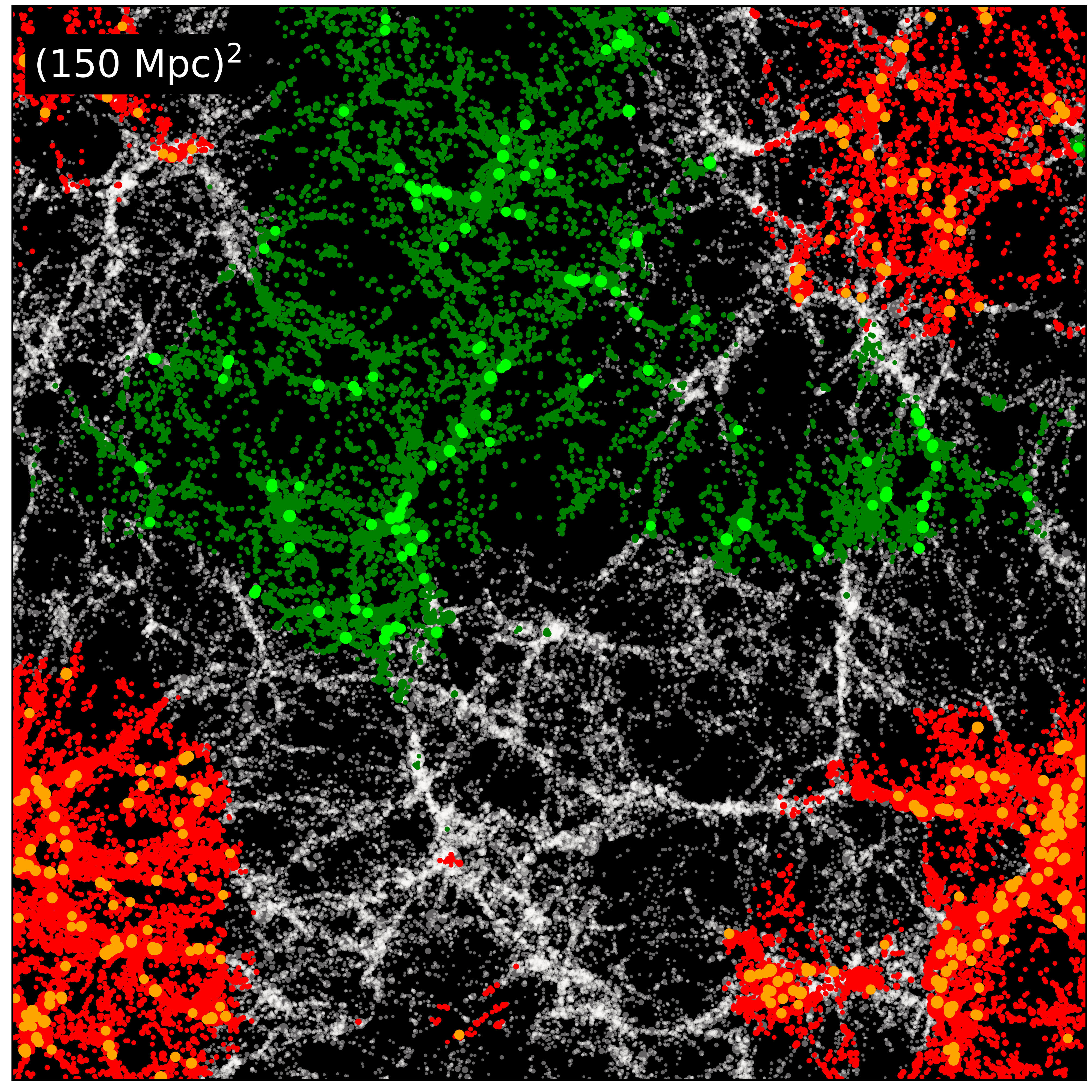}
        \includegraphics[width=.35\textwidth]{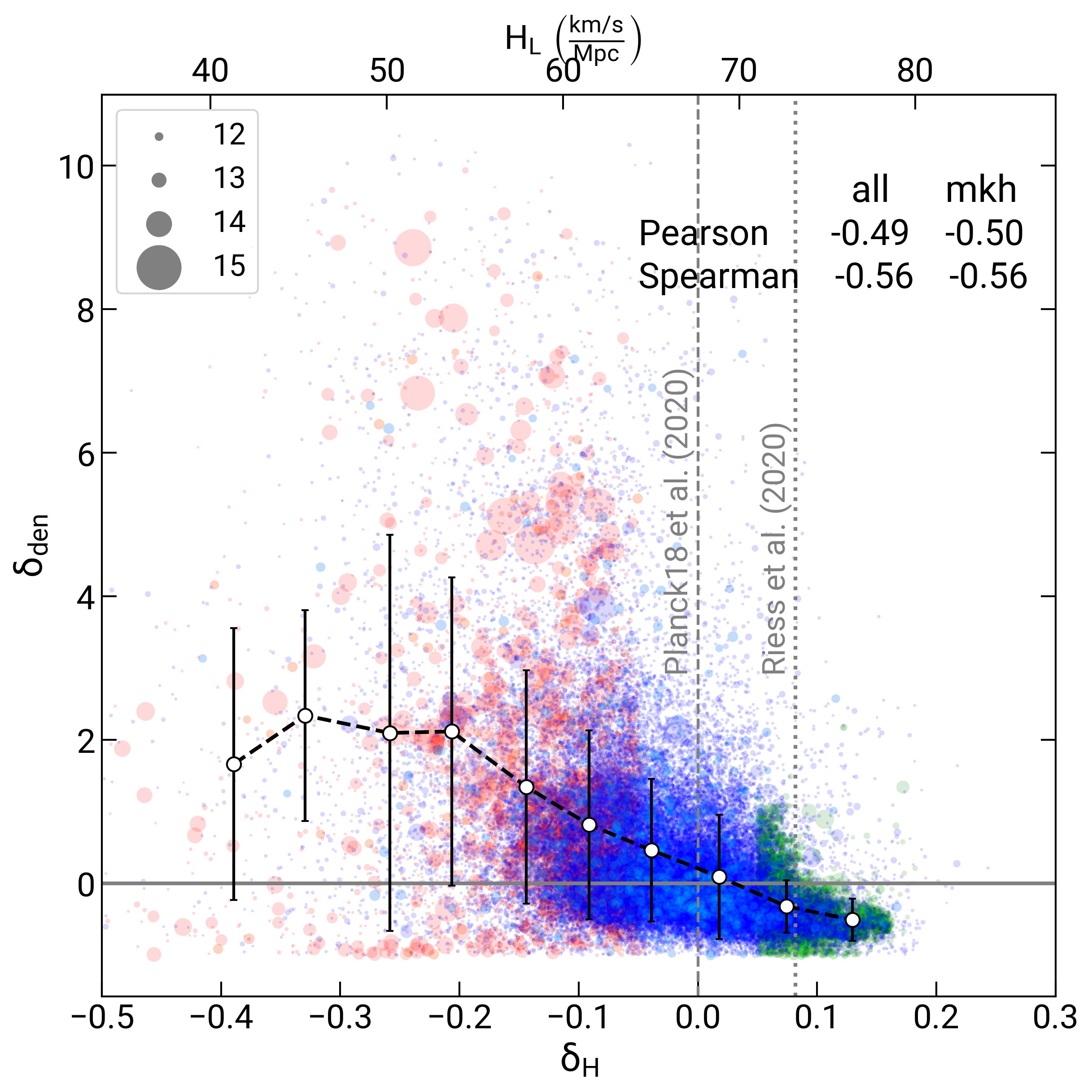} \\
        \includegraphics[width=.35\textwidth]{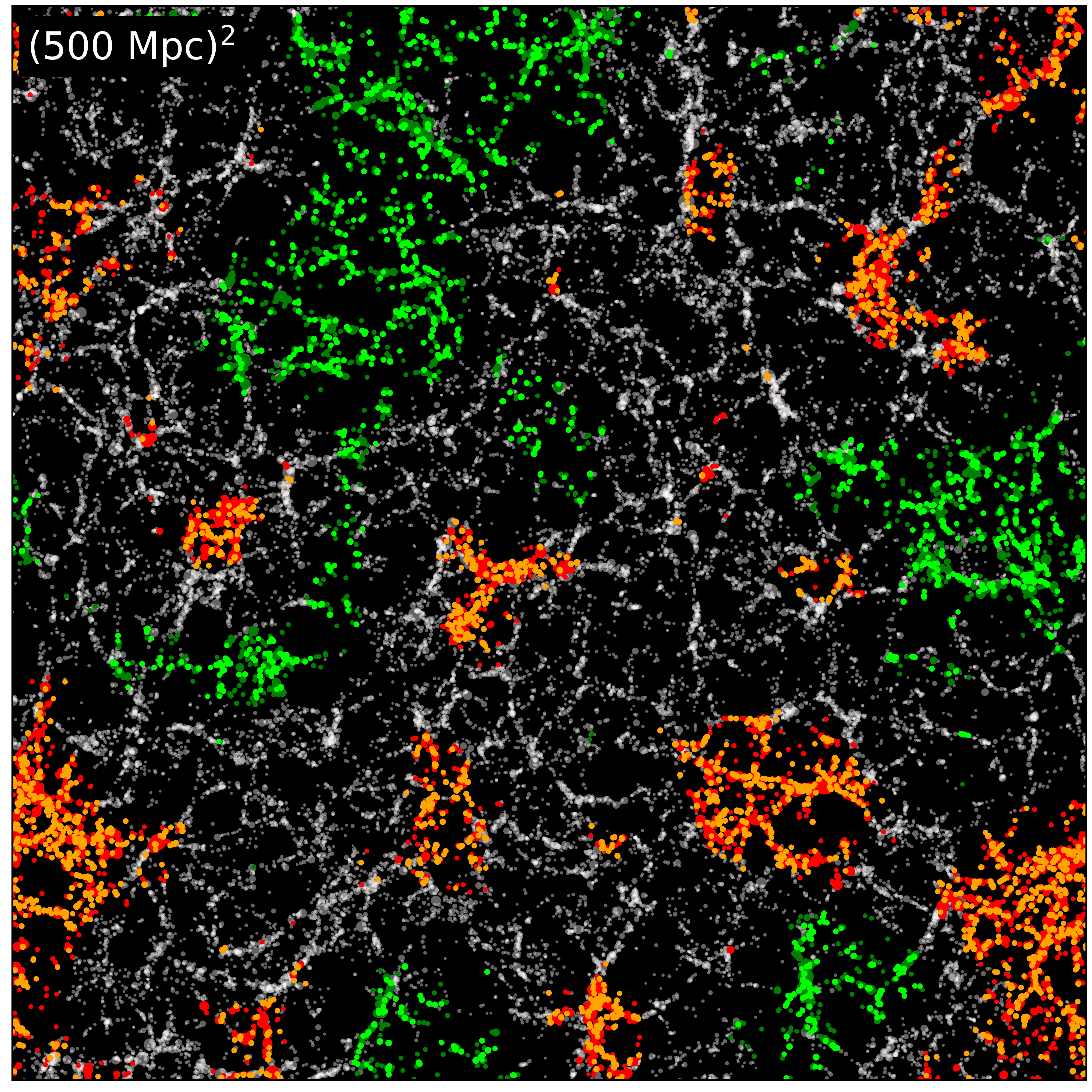}
        \includegraphics[width=.35\textwidth]{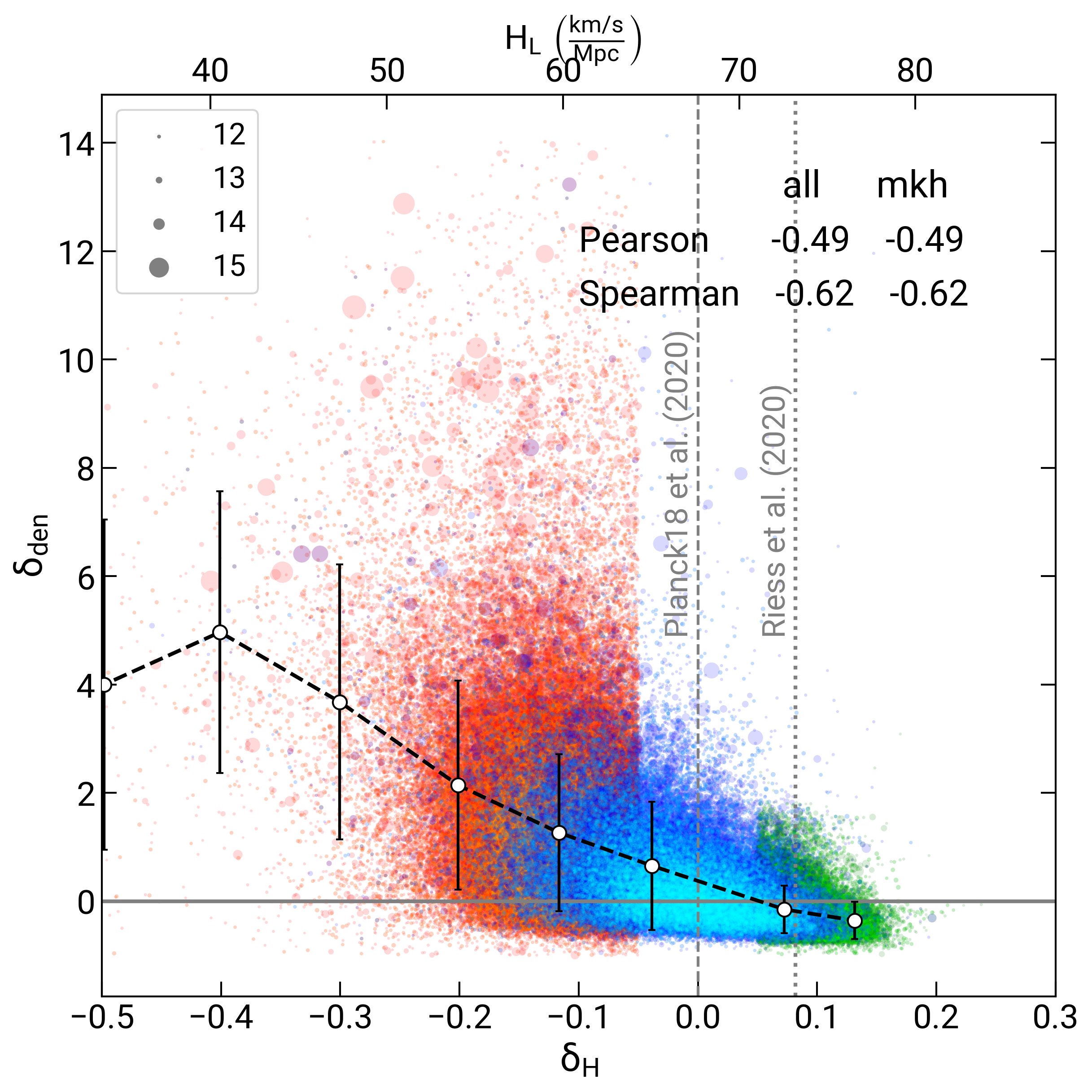} \\
        \includegraphics[width=.35\textwidth]{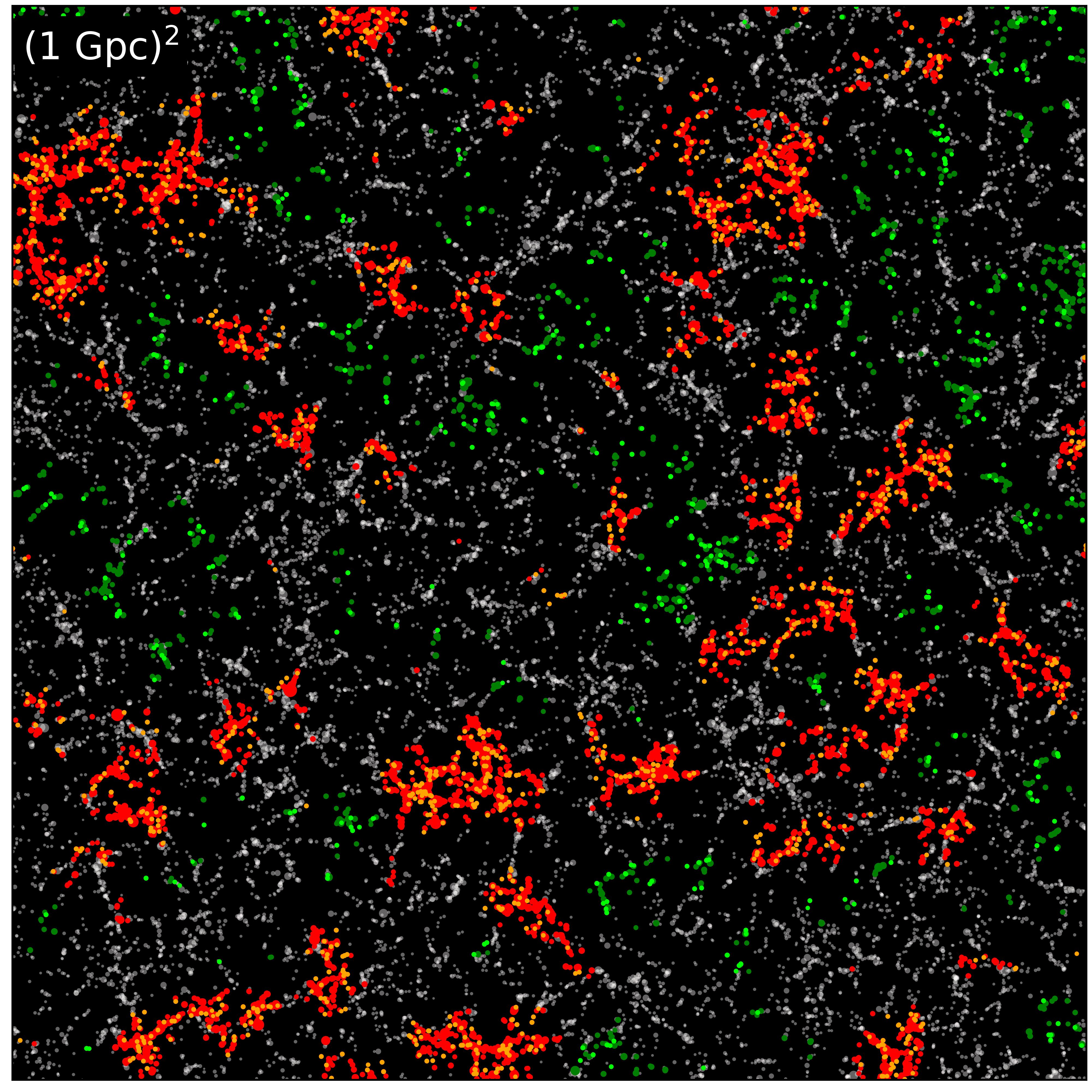}
        \includegraphics[width=.35\textwidth]{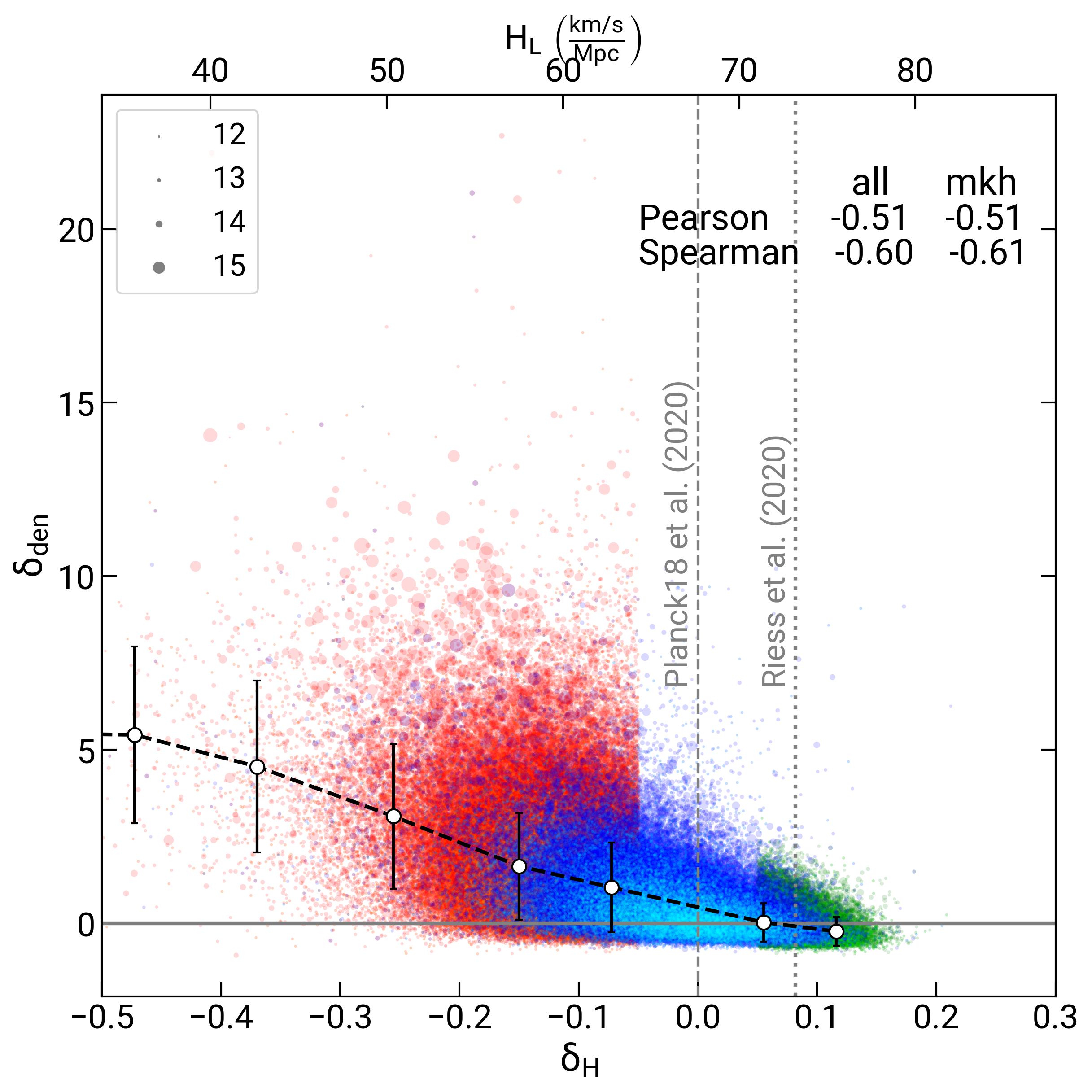}
    \caption{{\bf Correlation with local overdensity}. Left panel: A $10$ Mpc/h slice of simulation box showing all halos in blue color, red and green correspond to halos with deviations exceeding $5\%$. The color red (green) for halos with negative (positive) deviation with respect to the global Hubble value. The sizes of the halos are proportional to their masses, with the $\text{log}_{10}$ exponent indicated in the legend of the left panel. Right panel: The plot shows the local overdensity of the halos within $10$ Mpc/h on the y-axis against the deviation in Hubble constant measurement for it on the x-axis. The mean and standard deviation in bins for three scatters(red+green, red, green) are plotted over scatters. The solid black line represents the combined scatter, while the dashed black lines, with red and green markers, represent the individual red and green scatter. The Pearson and Spearman correlation coefficient is displayed for combined scatter (refer section \ref{ssec:corre}).}
    \label{fig:dev5p}
\end{figure*}

We identify halos that exhibit a deviation of more than $5\%$ in their distributions between $25$-$40$ Mpc/h. Approximately $10-15\%$ of all observers show this characteristic. To visually represent these halos, we have included $10$~Mpc/h thick slices of the simulation box in the left panels of figure \ref{fig:dev5p}. 
The scatter points in white denote all halos within the slice, while those in red and green correspond to halos with deviations exceeding $5\%$. We use the color red for halos with negative deviation from the global Hubble value and green for those with negative deviation. We use orange and light green colors to highlight Milky Way-sized halos, with mass ranging from $6 \times 10^{11}$ to $3 \times 10^{12} M_{\odot}$/h.
The sizes of the halo markers are proportional to the halo mass, with the $\text{log}_{10}$ exponent indicated in the legend of the right panels. 

We see regions with a lower than global Hubble constant, as determined above, lie in and around clusters, whereas regions with a higher determination are away from rich clusters. 
To quantify this, we examine the local surroundings of these identified halos to quantify their correlation. Specifically, we calculate the local over-density ($\delta_{\text{den}}$) of a halo within a $10$ Mpc/h radius and plot this against the deviation in Hubble constant measurement ($\delta_H$) for it within $25$-$40$ Mpc/h volume. The scatter from all halos is presented in the right panels of figure \ref{fig:dev5p}, with the same formatting as described for the left panels. We segment the data points for the all and Milky Way-sized halos into separate bins. We overplot the mean and standard deviation for each bin for the data sets. The dashed black line represents all the scatter. Further, we computed the Pearson and Spearman correlation coefficients for these, and the results are tabulated in the right panel of figure \ref{fig:dev5p}, where columns 'all' and 'mkh' represent all and  Milky Way-sized halos. We find a moderate negative correlation between $\delta_H$ (deviation in Hubble constant) and $\delta_{\text{den}}$ (local over-density).


\subsection{Milky Way-sized halos}
\label{ssec:mkh}
\begin{figure*}
    \centering
        \includegraphics[width=.4\textwidth]{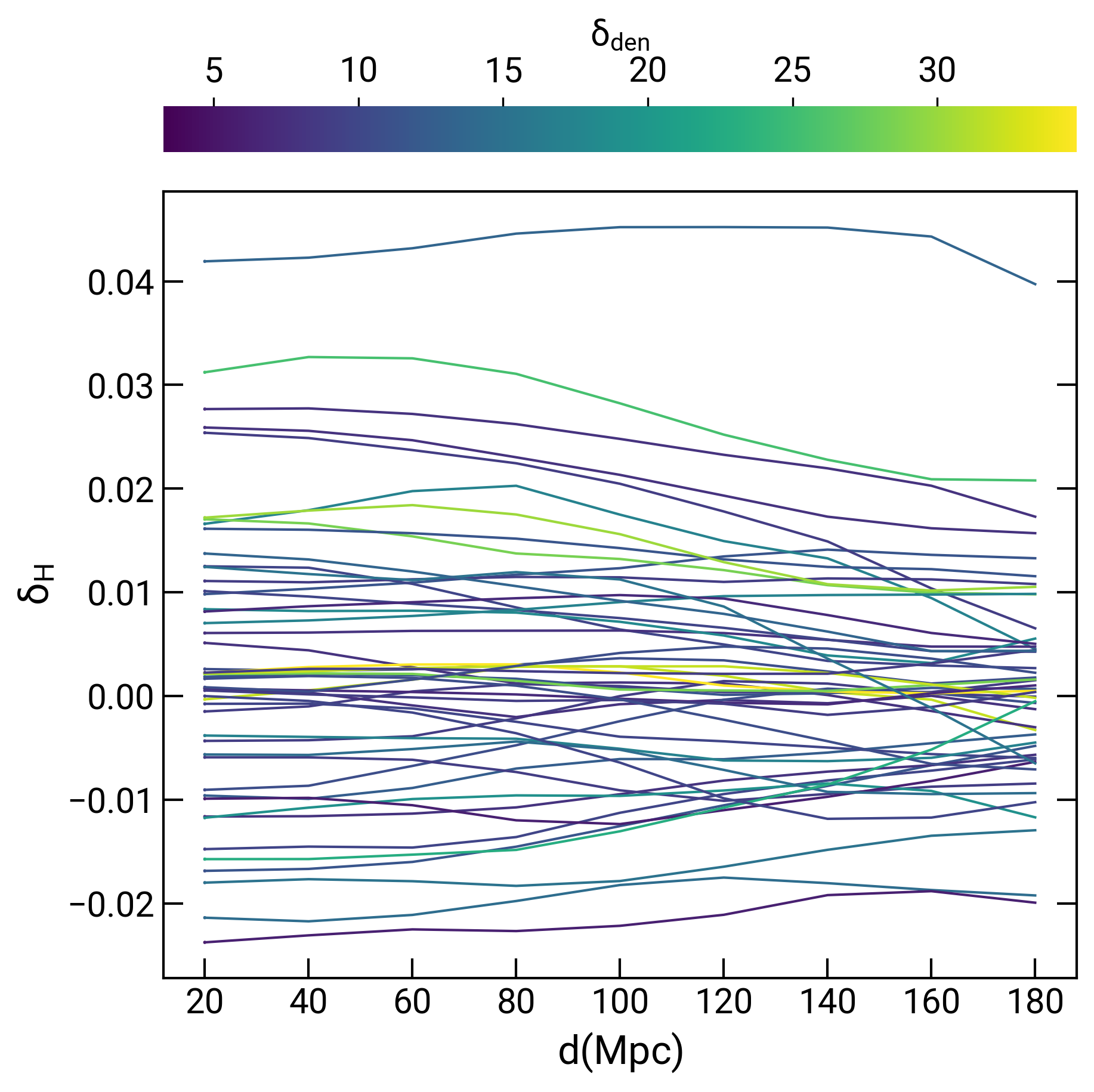}
        \includegraphics[width=.4\textwidth]{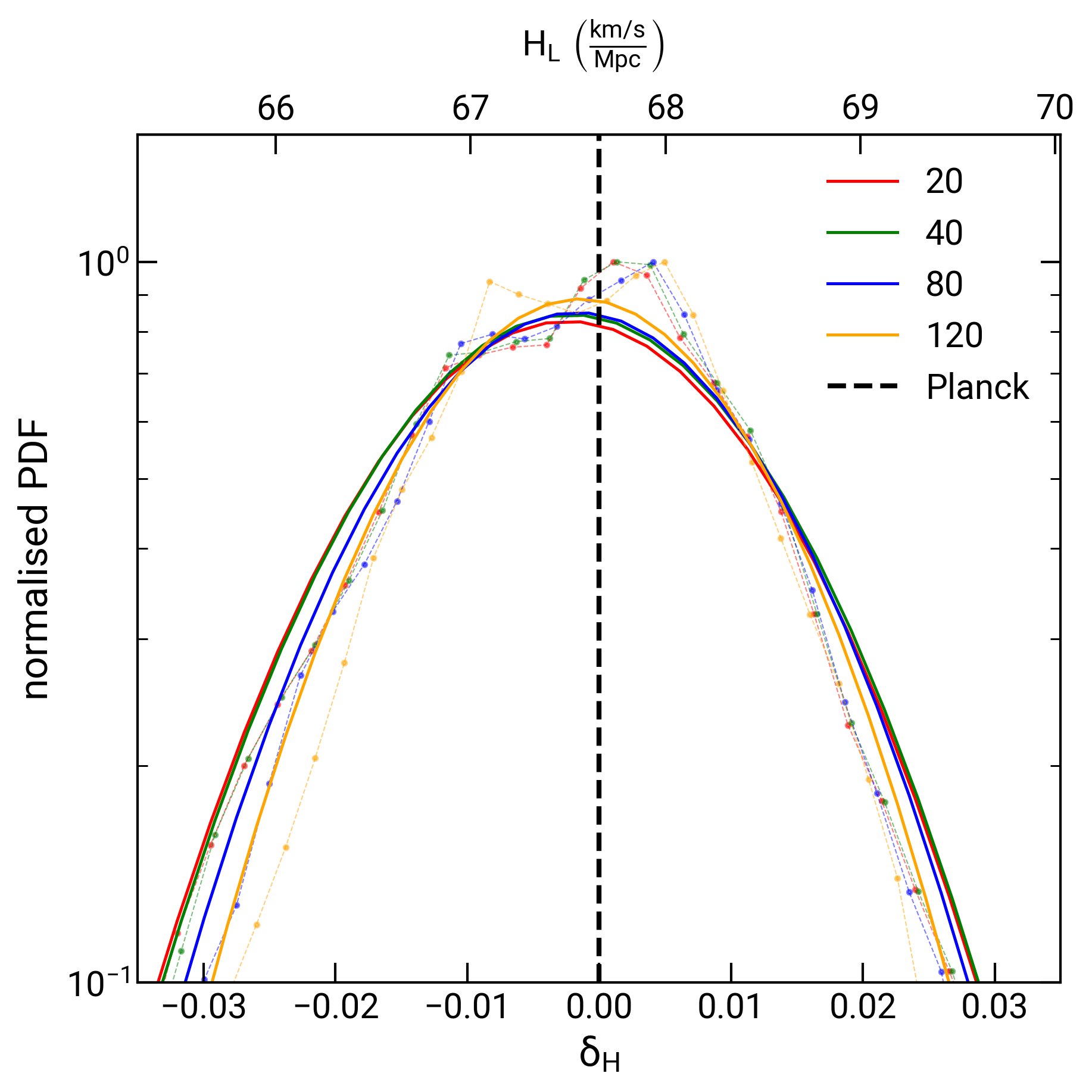}
    \caption{{\bf Milky way like halos:}.The left panel of figure \ref{fig:mkh} shows the deviations in local measurements for 50 randomly selected Milky Way-sized halos across various spherical shells with lower radii ($r_{\text{min}}$) shown on the x-axis. The upper radius ($r_{\text{max}}$) for all measurements is fixed at $200$ Mpc/h. The colors of the lines represent the local overdensity of these halos, as shown in the colorbar. The right panel show normalised PDFs for Milky-sized halos within spherical shells (20-200), (40-200), (80-200), and (120-200).}
    \label{fig:mkh}
\end{figure*}

\begin{table*}
  \fontsize{9pt}{9pt}\selectfont
  \begin{center}
    \begingroup
    \setlength{\tabcolsep}{9pt}
    \renewcommand*{\arraystretch}{1.7}
    
    \begin{tabular}{c c c c c}
      \hline 
      Shell($r_{\text{min}} - r_{\text{max}}$) & (20-200) & (40-200)  & (80-200) &  (120-200) \\
      \hline 
      1$\sigma$ & [-0.045, 0.037] & [-0.044, 0.037] & [-0.042, 0.038] & [-0.046, 0.036]\\ 
      \hline
      2$\sigma$ & [-0.042, 0.034] & [-0.042, 0.034] & [-0.040, 0.033] & [-0.039, 0.034]\\ 
      \hline
       3$\sigma$ & [-0.042, 0.034] & [-0.042, 0.034] & [-0.040, 0.033] & [-0.037, 0.032]\\ 
      \hline
      
    \end{tabular}
    \endgroup
    \caption{\label{tab:mkh}{\bf:1$\sigma$, 2$\sigma$ and 3$\sigma$ confidence intervals for PDFs shown in figure \ref{fig:mkh}}} 
  \end{center}
\end{table*}

The analysis presented above is limited to spherical shells.  
However, the observational determination of Hubble's constant relies on observations across a wide range of scales, often using multiple distance indicators.  
To emulate this, we randomly select 50 Milky Way-sized halos within our $500$ Mpc/h simulation box.  We ensure that these halos are not located close to each other to ensure the statistical independence of our analysis. 
The left panel of figure \ref{fig:mkh} depicts the deviations in local measurements across various spherical shells observed by these individual observers. On the y-axis we represent these deviations, and the x-axis denotes the lower radius ($r_{\text{min}}$) of the spherical shell. The upper radius ($r_{\text{max}}$) for all measurements is fixed at $200$ Mpc/h. We color the trajectories of these observers according to their local overdensity, as shown in the colorbar at the top.

In the right panel of figure \ref{fig:mkh}, using a $500$ Mpc/h simulation box, we show normalised PDFs for Milky-sized halos within spherical shells (20-200), (40-200), (80-200), and (120-200). Table \ref{tab:mkh}. 1$\sigma$, 2$\sigma$ and 3$\sigma$ intervals for these PDFs is tabulated in table \ref{tab:mkh}.
We note that the standard deviation of the distribution of $\delta_H$ is about a factor $2$ higher when compared to the stated standard deviation in SH0ES. We also note that a low probability tail extends to fairly large deviations from the global value.
 

\section{Conclusion and Discussion}

In this work, we have considered the effect of peculiar velocities on the error budget of the measurement of Hubble's constant in the local Universe.  
We have worked with a model where we consider observers and galaxies to be located in halos.  
The halo distribution and velocities are obtained from N-body simulations.  
We make a simplifying assumption of one observer and one target galaxy per halo.  
This is an oversimplification as it ignores subhalos, however this is expected to lead to a conservative estimates of the contribution of peculiar velocities as we ignore internal velocities within halos and the finger of god effect.  

Our key conclusions may be summarised as follows.
\begin{itemize}
\item 
We reproduce the result that the local measurements of the Hubble constant exhibit significant and systematic differences from the global value due to gravitational clustering and peculiar motions. 
\item 
We observe a negative correlation between the deviation of Hubble measurements from the global value and the local halo population. While earlier studies indicated such a correlation \cite{2014JCAP...10..028O,2014MNRAS.438.1805W,2017JCAP...03..022O}, this work provides a novel result by explicitly quantifying it.
\item 
The study suggests that, regardless of how accurately we measure the local value of the Hubble constant, it cannot be directly interpreted as a measure of the global value and deviations of the order of $3-5\%$ may be expected. 
\end{itemize}

We would like to note that observations have also revealed coherent bulk flow at scales in excess of $100$~Mpc/h in the region around the Milky Way galaxy.  
This puts our region as an outlier in the expected distribution of galaxies and flows in the $\Lambda$CDM model \cite{2023MNRAS.526.3051W, 2024MNRAS.527.3788H, 2024ApJ...967...47L}.  
An assessment of how these flows may contribute to the measurement of Hubble's constant requires improved simulations as we will need to simulate a much larger physical region than we have done so far while retaining high resolution.  
We plan to address this issue in a follow-up study while noting that if large-scale bulk flows indicate that we are away from the average, then it may very well be that we are in the tail of the distribution for observed $H_0$. 
Whether this can help alleviate the Hubble tension is something that can only be answered with more detailed studies. 

In summary, we would like to recommend that the impact of peculiar velocities and large scale structure needs to be studied in greater detail while interpreting results from observations.

\section*{Data Availability}
The {\tt SUBFIND} halo catalogues and {\tt GADGET4} snapshots of the simulations discussed in the text can be available upon request.

\section*{Acknowledgements}

Authors would like to thank Surhud More, Aseem Paranjape, Shadab Alam, Arif Babul, Pierluigi Monaco, and Stephane Colombi for useful discussions and suggestions.
JSB also thanks NCRA-TIFR for hospitality as this manuscript was
completed during sabbatical from IISER Mohali. 
This research has made use of NASA's Astrophysics Data System.

%

\end{document}